# Cross-platform modeling of users' behavior on social media

*Haiqian Gu[1,2], Jie Wang[2*], Ziwen Wang[1], Bojin Zhuang[2], Wenhao Bian[1], Fei Su[1]*

[1]*Beijing Univeristy of Posts and Telecommunications*

[2]*Ping An Technology (Shenzhen)*

*jayjay.wang@connect.polyu.hk*

*Abstract*—With the booming development and popularity of mobile applications, different verticals accumulate abundant data of user information and social behavior, which are spontaneous, genuine and diversified. However, each platform describes users' portrait in only certain aspect, and it is difficult to combine those internet footprints together. In our research, we proposed a modeling approach to model users' online behavior across different social media platforms. Users' data of same users shared by NetEase Music and Sina Weibo was collected for cross-platform modeling of correlations between music preference and other users' characteristics. Based on music genre and mood tags, users are clustered into five typical genre groups and four typical mood groups by analyzing their collected song lists. Moreover, based on collected Weibo data of same users, correlation between music preference (e.g. genre, mood) and Big Five personalities and basic information (e.g. gender, region, self-description tag) have been comprehensively studied, forming full-scale user profiles with finer grain. The results indicate that people's music preference can be reflected by their social activities in many ways. For example, people living in mountainous areas generally have favor for Folk music, while people living in rich regions tend to like Pop music. Meaningly, dog lovers prefer Sad music compared to cat lovers. Further discussion has been suggested in our paper. Promisingly, our proposed cross-platform modeling approach could be extended to other verticals, providing an online automatic way for profiling users more precisely and comprehensively.

## I. INTRODUCTION

With the rapid development of social media nowadays, people spend more time on different social media applications to share various aspects of their lives. For instance, they post their thoughts, feelings, or favorite songs or singers, leaving abundant electronic footprints online. For instance, Sina Weibo is a very popular social media in China analogy to Facebook and Twitter where 100 million microblogs are posted every day[2]. Since microblog text can reflect personality, much research on relationship between language style of posts and personality of social media users has been conducted. On the other hand, NetEase Music is one of the most popular music streaming providers in China. As indicated by official report of NetEase Music, listening to music is an frequent dailylife activity of people[1]. In this social media platform, varieties of songs including lots of different music genres, moods and languages have been shared. Moreover, users of this music platform can share their favorite song playlists through clicking a "Red Heart" button. Due to different function orientations, NetEase Music and Sina Weibo accumulate different user data but isolated from each other. Fortunately, these two platforms share parts of users with binding logging. Thus, basic user information and post texts from Sina Weibo can be used to supplement users' profiles of Newease Music by analyzing correlations between music preference and other user characteristics.

However, music preference and its potential influence on users' profile have been rarely exploited in the context of social networks. For instance, correlation modeling between music preference and online behaviors of social media users has not been comprehensively studied. To fill this gap, we propose a novel cross-platform modeling approach through data bridge between two difference social media platforms. In our paper, we first analyze music preference by use of cluster computing and then conduct correlation modeling with Big Five personality, basic information and interest tags collected from Sina Weibo. Our methodology and findings would contribute to optimizing online customer relationship management, such as full-scale user profile, precise product recommendation and personalized customer service and so on.

Key contributions of our work are summarized as follows and Fig. 1 shows key components of our system:

- **Cross-platform modeling approach of user profiles by use of data bridge between two different main social media platforms**. Through data crawling from NetEase Music and Sina Weibo, cross-modal user data including structural and non-structural data are fused for social media computing.

- **Modeling of music preference by cluster computing of song lists.** Divided user groups show discriminative and independent music preference, providing valuable and meaningful basis for further analysis of user profile.

- **Correlation analysis between music preference and typical behaviors on social media of users has been comprehensively conducted.** We analyze correlation between music preference and Big Five personality, sex, age, zodiac and geographic region of users with reasonable explanations.

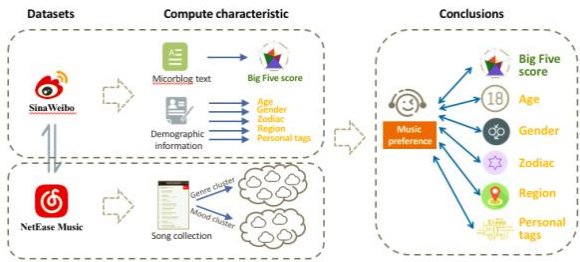

Fig. 1. Key components of our system.

## II. RELATED WORK

Previous researches suggest that musical preference is linked to individuals personality and personal identities. Rentfrow et al.[4] concluded that people used preferred music to convey their personalities to others. Little and Zuckerman [5] noted that preferred music style of people could reinforce and reflect their personalities. In addition, Rentfrow et al., [6] claimed that music preference was also related to a wide variety of personality aspects, self-views (e.g., political orientation), and cognitive abilities (e.g., verbal IQ). In fact, when individuals expressing music preferences, they were essentially delivering that they shared similar attitudes, values and beliefs with the other members of their group [7].

On the other hand, several modeling methods of personality have been studied for user portrait based on user information and social behavior. Mehl et al.[8] found that Big Five personality of internet users were correlated with their daily life behaviors, including the way of language use, daily social interactions, locations, activities, and moods. Furthermore, Golbeck et al.[9] demonstrated that public information shared on Facebook could be used to predict users' Big Five personality. Additionally, Hung et al. [10] put forward a tag-based user profiling method for social media recommendation. In a word, music preference, daily activities, demographic information and other social media network behaviors of internet users can be fused for a full-scale user profile.

## III. METHODOLOGY

### A. Data Preprocessing

Firstly, a seed account was used to crawl its followers and corresponding favorite song lists, scaling up to 14,427 users. With the help of superlink between NetEase Music and Sina Weibo, we selected those shared users and prepared their public information for further analysis. In this way, favorite song lists of Netease Music have been collected including all detailed songs. Through superlinked Sina Weibo IDs, we then collected their recent 200 microblogs and basic information including age, gender, location, self-description tags. Additionally, we have used 111 music labels for music classification including genre, mood and scene. In this paper, genre and mood labels could be used for analysis of music preference.

Furthermore, valid users are defined who have collected more than 20 songs in Netease Music and posted more than 500 valid texts in Sina Weibo. Eventually, totally 11,860 active NetEase Music users and 9,442 active Sina Weibo users have been filtered out for further explorations of users' profiles.

### B. Modeling and Analysis

In this paper, we modeled users' music preference by using K-means clustering and their collected song lists. The formed user groups show discriminative music preference, providing valuable and meaningful basis for further analysis of user profile.

The Big Five Personality has been widely used for analysis of social media users, including five personality traits, i.e. Openness, Conscientiousness, Extraversion, Agreeableness and Neuroticism. Previous work has proved that linguistic styles can be used to compute users' personality profiles without customer disruption. Based on LIWC linguistic features[11], BFP scores have been calculated for all valid users.

The BFP is characterized as shown in TABLE I. [12]. As an important psychometric method proved by many researchers, BFP modelling has been proved for obtaining comprehensive profiles of users' cognitive patterns.

## IV. CLUSTERING RESULTS OF MUSIC PREFERENCE

### A. Music Preference Cluster of Genre

In this research, 13 music genre tags have been examined, including Ancient Chinese, Country, Classical, Downtempo, Rock, New Age, Folk, Pop, Jazz, Electronic, Blues, Light music, HIP-HOP. More importantly, a song genre mapping has been prepared based on music streaming media tags. In this way, a music genre preference vector has been formed for all valid users based on their collected song lists. Later, K-means algorithm has been used to divide all users into five clusters as follows:

1) Based on favourite song list of each user, genre tags have been counted to obtain a 13-dimention music genre preference vector.

2) The The music genre preference vectors of all users have been normalized and then clustered by means of K-means method. Thus, Ancient Chinese Style (ACS) Cluster, Rock Cluster, Pop Cluster, and Folk Cluster and Neutral Cluster are obtained.

TABLE II. shows the detailed result of cluster centers and the percentage of each cluster. From the table, we can see the first cluster center G1 shows an obvious preference for Ancient Chinese Style(33%) music, named ACS Cluster and the second center G2 prefers Rock(18%) and Blues(15%) music, named Rock Cluster. The third cluster center G3 is keen on Pop(61%) music, named Pop Cluster, and the fourth cluster center G4 loves Folk(43%) music, named Folk Cluster. However, the fifth cluster center G5 shows no apparent music genre preference compared with the rest, named Neutral Cluster.

TABLE I. THE DEMENSIONS OF BIG FIVE PERSONALITY

| Dimension | Score | Personal traits |
|---|---|---|
| Openness | high | Wide interests, Imaginative, Intelligent, Curious |
| | low | Commonplace, Simple, Shallow, Unintelligent |
| Conscientiousness | high | Organized, Tend to plan, Efficient, Responsible |

| | low | Careless, Disorderly, Frivolous, Irresponsible |
|---|---|---|
| Extroversion | high | Talkative, Active, Energetic, Enthusiastic |
| | low | Quiet, Reserved, Shy, Silent |
| Agreeableness | high | Sympathetic, Kind, Appreciative, Generous |
| | low | Fault-finding, Cold, Unfriendly, Cruel |
| Neuroticism | high | Tense, Anxious, Nervous, Worried |
| | low | Stable, Calm, Contented, Unemotional |

TABLE II. CENTROIDS AND DISTRIBUTION OF PREFERENCE CLUSTERS OF MUSIC GENRE

| Genre | G1 | G2 | G3 | G4 | G5 |
|---|---|---|---|---|---|
| Ancient Chinese Style | **0.33** | 0.03 | 0.07 | 0.04 | 0.17 |
| Rock | 0.05 | **0.18** | 0.05 | 0.15 | 0.10 |
| Folk | 0.10 | 0.08 | 0.11 | **0.43** | 0.19 |
| Pop | 0.29 | 0.22 | **0.61** | 0.23 | 0.30 |
| Blues | 0.04 | **0.15** | 0.04 | 0.03 | 0.07 |
| Cluster name | ACS | Rock | Pop | Folk | Neutral |
| Clutser Percentage | 5.3% | 20.4% | 15.1% | 16.2% | 43.1% |

TABLE III. CENTROIDS AND DISTRIBUTION OF PREFERENCE CLUSTERS OF MUSIC MOOD

| Mood | M1 | M2 | M3 | M4 |
|---|---|---|---|---|
| Sad | **0.10** | 0.07 | 0.10 | 0.02 |
| Lonely | **0.10** | 0.07 | 0.10 | 0.03 |
| Quiet | 0.09 | 0.09 | **0.16** | 0.04 |
| Joyful | 0.05 | 0.09 | 0.04 | **0.22** |
| Curing | 0.09 | 0.09 | **0.13** | 0.08 |
| Cluster name | Sad | Neutral | Quiet | Joyful |
| Clutser Percentage | 58.9% | 22.1% | 12.8% | 6.2% |

*B. Music Preference Clustering of Mood*

In a similar way, analysis of music mood preference has been conducted by using 22 music mood tags including Sad, Inspiring, Rebelling, Lonely, Quiet, Vent, Missing, Perceptual, Joyful, Curing, Sweet, Confusing, Calm, Heartbroken, Frustrating, Expecting, Relaxing, Vulnerable, Romantic, Happy, Cherish, Distressing forming four clusters as follows:

1) Count music mood labels based on users' collections of favourite songs and calculate a 22-dimention music mood preference vector for each user.

2) Normalize music mood preference feature vector and carry out K-means based clustering. Thus, Sad Cluster, Neutral Cluster, Quiet Cluster, and Joyful Cluster are obtained.

TABLE III. shows the detailed result of cluster centers and the percentage of each cluster. From the table, we can see the first cluster center M1 prefers Sad(10%) and Lonely(10%) music, named Sad Cluster and the second center M2 shows no apparent music mood preference, named Neutral Cluster. The third cluster center M3 likes Quiet(16%) and Curing(13%) music, named Quiet Cluster, and the fourth cluster center M4 has an obvious preference for Joyful(22%) music, named Joyful Cluster. Divided user groups show discriminative and independent music preference. Every user has been clustered into one music genre preference cluster and one music mood preference cluster, which providing a valuable and meaningful basis for further analysis of user portrait.

*C. Crosscorrelation between Two Clusterings*

Firstly, we have studied crosscorrelations between the two clusterings in previous sessions. As shown in Fig. 2, clusters of Ancient Chinese Style music and Pop music mainly compromise of Sad music, while Rock music cluster mainly consists of Joyful music. Additionally, the folk cluster is closely correlated with Quiet music.

Conversely, music of different mood preferably contain music of different genres. As indicated in Fig. 3, Sad music is deminantly linked with Pop music, and people who prefer Quiet music almost choose Folk music. In addition, people who like Joyful music will probably like Rock music.

Therefore, mood and genre types of music can be correlated with each other. For instance, most Pop music is classified as Sad music; most Rock music is considered as Joyful, while Folk music is recognized as Quiet music.

V. ANALYSIS OF MUSIC PREFERENCE CLUSTERING RESULTS

*A. Gender and Music Preference*

Gender information of users can be inferred from Weibo profiles including 37.4% female and 62.6% male ones.

As shown in Fig. 4, male users prefer Rock music, while female users like Ancient Chinese Style and Folk music more. As for music mood, male users are funds of joyful music while female users prefer Sad music, since female users are more sensitive with rich affection than male ones.

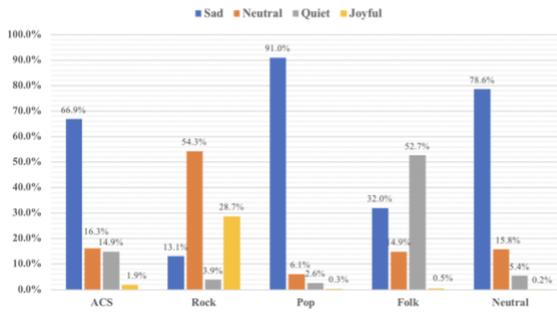

Fig. 2. Distribution of four music-mood clusters in all five music-genre groups.

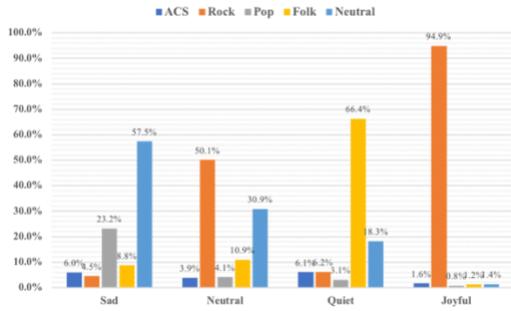

Fig. 3. Distribution of five music-genre clusters in all four music-mood groups.

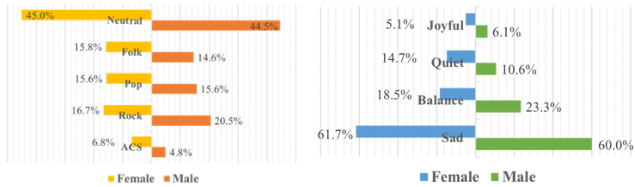

Fig. 4. Music preference of genre (left) and mood (right) dependent on gender.

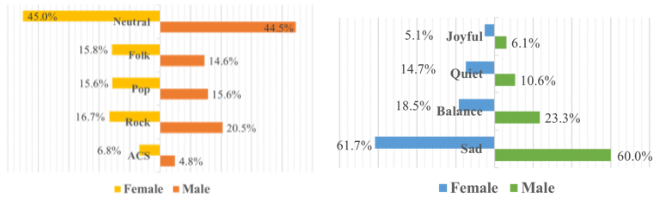

Fig. 5. Music preference of genre (left) and mood (right) dependent on age

### B. Age and Music Preference

Based on birthday information of users shared in Sina Weibo, users are devided into two special groups: post-90s and post-80s. Among them, post-90s are born between 1990 and 1999, while post-80s are those born between 1980 and 1989. As shown in **Error! Reference source not found.**, it could be concluded that post-90s prefer Pop music and Sad music, while post-80s prefer Folk music and Joyful music.

### C. Zodiac and Music Preference

Correlation between Zodiac and users' music preference has also been calculated. According to our study, Capricorn and Taurus are fond of Rock music while Pisces are not very keen on Pop music. Most of Aquarius users belong to Neutral music genre cluster while most of Taurus users prefer Rock or Folk music. Music preference of genre among zodiacs is depicted in Fig. 6 .

In terms of music preference of mood, Aquarius generally like Sad music, while Taurus generally dislike Sad music. Most of Aries users belong to the cluster of Neutral music preference of mood and prefer Quiet music. Few of Sagittarius users prefer Neutral preference while exhibit an extreme preference of music mood For instance, most of Sagittarius users like either Sad or Joyful music, and they hate Quiet music. Music preference of mood among zodiacs is plotted in. Fig. 7.

### D. Region and Music Preference

In our study, region of user has been proved as one of the most significant demographical factors for understanding users' music preference. In China, six main areas have been used for further study based on geographical relation, including North China, Northeast China, East China, South Central China, Southwest China, Northwest China. TABLE IV. shows detailed information about the six main areas of China, including economic level, topographic features, inhabitation of ethnic minorities and so on.

As shown in Fig. 8, it can be obsered that people from different regions prefer different music genres. For instance, people from East and South Central China prefer Pop music, while those from South Central China like Ancient Chinese Music. People from Northwest China are keen on Rock music while those from Southwest China tend to love Folk music. Taking into condiseration terrain and economic factors, it can be concluded as follows:

1) Preference of pop music is closely related to economic development level. For instance, users from prosperous areas are more open and have more chance to access Pop music than those from underdeveloped regions.

2) People from mountainous areas prefer Folk music. Since most Folk music sing the praises of simple and natural rural life, it can arouse sympathy of residents in mountain areas. As proved by Fox and Wince[13], inhabitants from suburban and rural regions tend to like classic rock, country, folk and oldies.

3) Some ethnics who like singing and dancing could prefer Rock music, since rhythmic music has become one of favorite entertainments of their daily life.

As shown in Fig. 9, people from different regions have different music preference of mood. For instance, people from Southwest China like Quiet music, while those from Northwest China could prefer Joyful music. Oppositely, people from Northwest China prefer Joyful music. Similarly, it can be concluded as follows:

1) People from high economic level regions more probably choose sad music than those from underdeveloped.

2) People's music genre preference can affect music mood preference, since the main mood for each region is mostly the main mood for their favorite music genre.

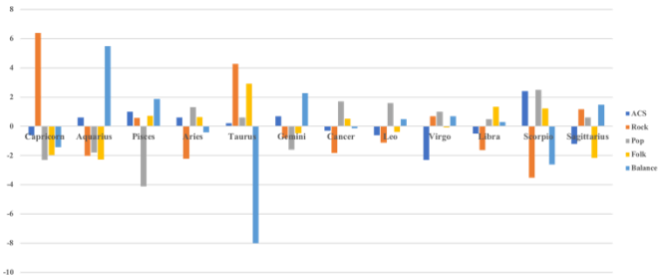

Fig. 6. Distribution of music preference of genre among different zodiacs.

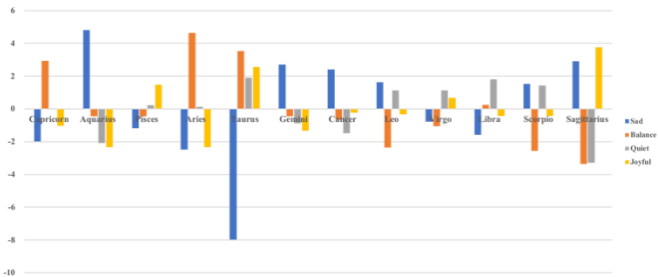

Fig. 7. Distribution of music preference of mood among different zodiacs.

TABLE IV. BSIAC GEOGRAPHIC AND ECONOMIC INTRODUCTION ON CHINA REGIONS OF WEIBO

| Region | Map | Per Capita GDP | Distinguishing Features |
|---|---|---|---|
| North China | | 8.124 | Rich and is the political, military, and cultural center of China |
| Northeast China | | 5.107 | Used to be rich |
| East China | | 8.979 | The richest region and covers the eastern coastal area of China |
| South Central China | | 5.532 | Mountainous and has rapid development |
| Southwest China | | 4.424 | The poorest region in China and has many mountains, also lived some Ethnic minorities |
| Northwest China | | 4.562 | Inhabited by many ethnic minority people |

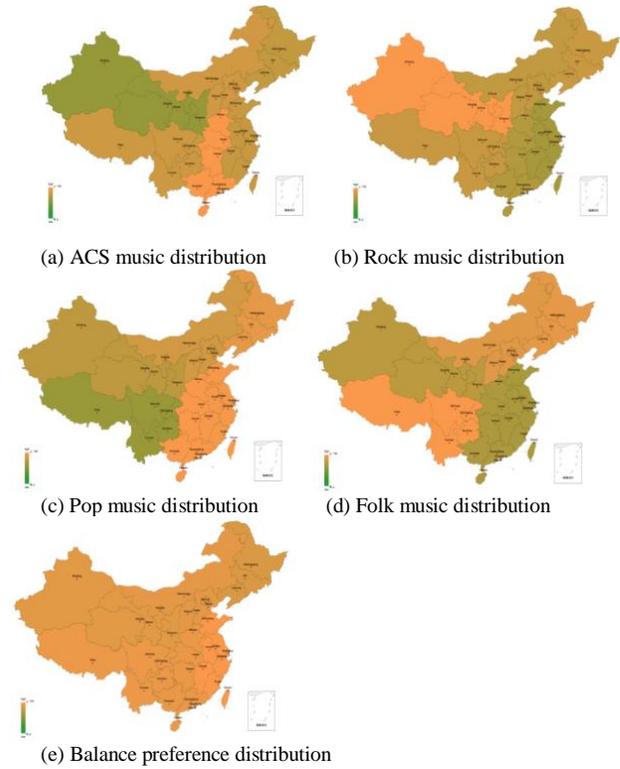

(a) ACS music distribution  (b) Rock music distribution

(c) Pop music distribution  (d) Folk music distribution

(e) Balance preference distribution

Fig. 8. The geographical distribution of different music preference of genre.

VI. MUSIC PREFERENCE AND BIG FIVE PERSONALITY

A. Preference of Music and Personality

As for each factor of Big Five personality, users from two polarity groups are selected to study correlation between BFP and music preference, including high-BFP group (i.e. the highest 30%) and low-BFP group (i.e. the lowest 30%). Fig. 10 shows distribution of high-BFP users in all preference clusters of music genre where vertical axis represents proportion of high-polarity users of corresponding BFP dimension with a bias of 50%.

As shown in the histogram of Fig. 10, people with different preference of music genre owns discriminative BFP traits. For example, people who are fond of Folk music get low scores in all five dimensions of BFP while those who prefer neutral music preference could get high scores in all five dimensions of BFP.

Fig. 11 depicts distribution of high-BFP users among all preference clusters of music mood. As shown in the histogram, people who are fond of Quiet music get low score in all five dimensions of BFP while those who are fond of Joyful music get high score in all five dimensions of BFP. Conversely, users with low scores in all five dimensions of BFP could be expected to own preference of Quiet music, while those with high BFP scores could prefer Joyful music. We notice that people love Joyful music usually score very high in Agreeableness, indicating that this group tend to be kind and have a good temper. From the two figures, we come to the conclusion that music preference can be a latent reflection of users' personality.

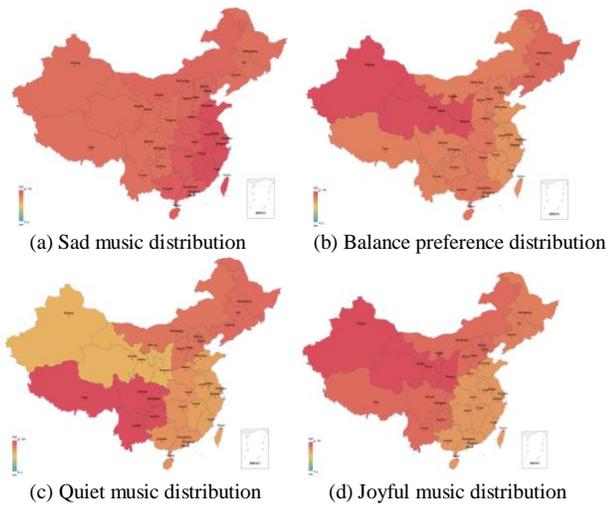

(a) Sad music distribution  (b) Balance preference distribution

(c) Quiet music distribution  (d) Joyful music distribution

Fig. 9. The geographical distribution of different music preference of mood..

## VII. CORRELATIONS BETWEEN MUSIC PREFERENCE AND SINA WEIBO TAGS

### A. Sina Weibo Tags

In Sina Weibo, users could write down up to 10 tags to share their interest, characteristics or profession. Among a variety of user-defined tags, those most frequent tags are selected for analysis of users' interest. Additionally, tags of same meanings are manually combined. For instance, tags of "Music", "Loving music" or "Music Fan" are merged into "Music" tag.

It is noteworthy that users' tags could be divided into three high-level categories: hobby, interest and character. Among them, hobby, interest and characteristics tags describe what users like to do, what users are interested in and who they are, respectively. For example, "Basketball" is a hobby tag, "Music" is an interest tag while "Humorous" is a characteristic tag.

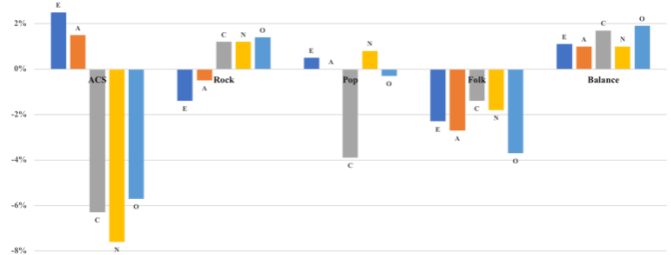

Fig. 10. Proportion of high personality in music genre preference cluster.

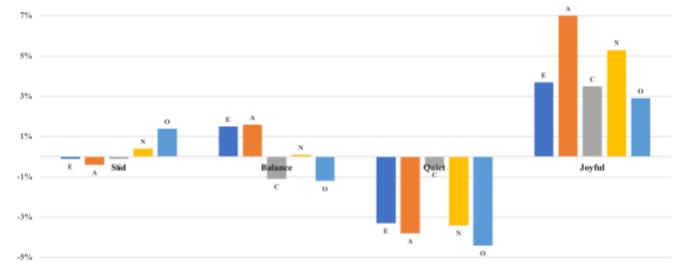

Fig. 11. Proportion of high personality in music mood preference clusters.

TABLE V. DIFFERENT MUSIC PREFERENCE AND SINA WEIBO TAGS

| Cluster | ACS | Rock | Pop | Folk | Balance |
|---|---|---|---|---|---|
| Hobby | Cosplay, Drawing, Physical Exercise, Mountain climbing, Bodybuilding, Badminton, ... | Dota, Playing, Swimming, Tennis, Football, Make Friends, ... | Entrepreneurship, Running, Basketball, Guitar, Shopping, ... | Dota, Sleeping, Tennis, Guitar, Playing, Mountain climbing, ... | Make Friends, Mountain climbing, Playing, Tennis, Swimming, Dota, ... |
| Interest | Architecture, Jay Chou, Classical Music, World of Warcraft, Fantasy, Comic, Piano, ... | ACG, History, Programmer, Philosophy, Rock, Mayday, Dog, ... | Beauty, Weibo, We Media, Japanese TV, Gossip, English, Pets, ... | We Media, Love, Jay Chou, English, Mayday, One Piece, Piano, ... | Mayday, Jay Chou, English, Psychology, We Media, Love, Piano, One Piece, ... |
| Character | Dreamy, Nostalgic, Sensitive, Embarrassed, Quiet, Bright, ... | Equanimity, Nervous, Fresh, Humorous, Kindhearted, Good-looks Club, Good Temper, Dreamy, Hardcore | Fighting, Kindhearted, Good Temper, Nervous, Good-looks Club, Love Laughing, Paradox, Optimistic, Quiet, Youth | Hardcore, Simple, Fighting, Dream, Daydreaming, Humble, Sunny, ... | Hardcore, Dream, Fresh, Fighting, Youth, Dreamy, Humble, Quiet, Perfectionism, Simple, Obsession |

### B. Music Genre Preference and Sina Weibo Tags

TABLE V. shows top tags of Weibo of all preference clusters of music genre. As shown in the table, users' Weibo tags are consistent with preference of music genre. For instance, users fond of Folk music usually own Simple, Sunny, Optimistic and Dreams tags. Some users who like Guitar could be interested in folk music. With help of Sina Weibo tags, more specific and detailed user portraits could be achieved for further analysis of music preference.

### C. Pet Preference and Music Preference

As a demonstration of tag analysis, two special type of tags of pets are discussed in detail. Two main groups of pet keeper are formed, that is, users who love cats and dogs. Limited by dataset size, 117 users who love cats and 40 users who love dogs are collected. As shown in Fig. 12, music preference of cats lovers is similar to that of normal users. Surprisingly, dog lovers prefer Sad music, and none of them lovers prefer Joyful or Quiet music!

It seems that cat lovers seek affection, while dog lovers look for companionship. Perhaps some people select pets based on their own personality, as noted ny Guastello[14]. It was claimed that cats are often considered independent animals while dogs are more obedient and warm. However, both cat and dog lovers prefer more Sad music than normal users, since users keeping pets are more lonely and sentimental.

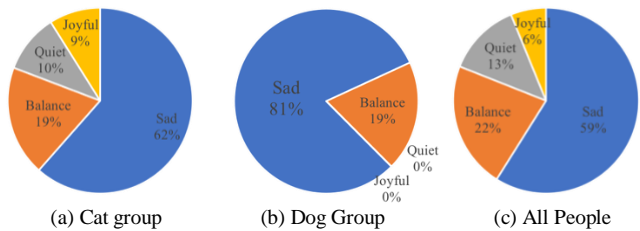

(a) Cat group  (b) Dog Group  (c) All People

Fig. 12. The difference of music mood preference between cat lovers and dog lovers.

## VIII. LIMITATIONS

The present research is subject to two limitations considering the generalizability of the finding results. One limitation is based on the fact that users on NetEase Music and Sina Weibo are almost young people, thus whether the findings would generalize to other age groups is worth further discussion.

Another limitation is that the studies were conducted in two mainstream social media platforms in China. Obviously, living environment and national culture can affect music preferences – individuals living abroad are exposed to different musical cultures compared with individuals living in China. Therefore, our research on music preference may not generalize to other geographic locations. Nevertheless, we hope that our findings can be of reference value for future research on music preference in other countries.

## IX. CONCLUSIONS AND FUTURE WORK

A cross-platform modeling of user portraits based on fusion of different social media data has been demonstrated, covering relationship between users' music preference and BFP traits, sex, age, zodiac and resident region. Publicly available data from two popular social networks in China has been collected and prepared for analysis of users. Specifically, K-means clustering method has been used to model users' favorite song lists on NetEase Music. Additionally, basic information of shared users from Sina Weibo has also been crawled to supplement users' portraits. The built user's portraits of music preference will be promising for precise marketing, commodity recommendation and optimal customer service.

As a part of our future work, cross-media users' data from cross-platform social media will be fused into our proposed model except those presented structral data. For instance, audio files of songs and vedios shared by users could be included for full-scale analysis of users and to build up a complete user's portraits for commercial applications.